\documentclass[onecolumn,floatfix,preprintnumbers, 11pt]{revtex4}

\usepackage{setspace}
\usepackage{graphics,amssymb,amsmath,epsfig,color}
\usepackage{graphicx}

\usepackage[raggedright, small]{titlesec}
\usepackage{ragged2e}
\usepackage{subcaption}
\DeclareCaptionJustification{justified}{\justifying}
\makeatletter

\def\@hangfrom@section#1#2#3{\@hangfrom{#1#2}#3}
\def\@hangfroms@section#1#2{#1#2}

\begin{document}


\title{Modulation-free Laser Stabilization Technique Using Integrated Cavity-Coupled Mach-Zehnder Interferometer}

\author{Mohamad Hossein Idjadi}
 \email{mohamad.idjadi@nokia-bell-labs.com}
\author{Kwangwoong Kim}%
 
\affiliation{%
 Nokia Bell Labs, 600 Mountain Ave, Murray Hill, NJ 07974, USA.
}%


\begin{abstract}
Stable narrow-linewidth light sources play a significant role in many precision optical systems. Electro-optic laser frequency stabilization systems, such as the well-known Pound-Drever-Hall (PDH) technique, have been key components of stable laser systems for decades. These control loops utilize an optical frequency noise discriminator (OFND) to measure frequency noise and convert it to an electronic servo signal. Despite their excellent performance, there has been a trade-off between complexity, scalability, power consumption, and noise measurement sensitivity. Here, we propose and experimentally demonstrate a modulation-free laser stabilization technique using an integrated cavity-coupled Mach-Zehnder interferometer (MZI) as an OFND. The proposed architecture maintains the sensitivity and performance of the PDH architecture without the need for any modulation. This significantly improves overall power consumption, simplifies the architecture, and makes it easier to miniaturize into an integrated photonic platform. An on-chip microring resonator with a loaded quality factor of 2.5 million is used as the frequency reference. The implemented chip suppresses the frequency noise of a semiconductor laser by 4 orders of magnitude. The integral linewidth of the free-running laser is suppressed from 6.1 MHz to 695 KHz. The passive implemented photonic integrated circuit occupies an area of 0.456 mm$^2$ and is integrated on AIM Photonics 180 nm silicon-on-insulator process. 
\end{abstract}

\maketitle

\section{Introduction}\label{sec1_intro}
Precise laser frequency control is a crucial requirement in various applications, including optical communication \cite{al2014ultra, blumenthal2020frequency}, optical atomic clocks \cite{ludlow2015optical}, microwave photonic \cite{li2013microwave}, and sensing \cite{marra2018ultrastable} which makes stable and narrow linewidth lasers indispensable part of precision optical experiments. Researchers have explored various methods to suppress unwanted laser frequency noise using optical feedback\cite{liang2015ultralow, dahmani1987frequency}, electro-optic feedback \cite{drever1983laser, idjadi2020nanophotonic}, and electro-optic feed-forward \cite{aflatouni2012wideband} techniques. The Optical feedback method relies on the Rayleigh scattering \cite{gorodetsky2000rayleigh} within the cavity and can be an effective and promising way to suppress the laser frequency noise. This mandates the co-design of laser system and optical signals with an ultra-high quality factor (Q-factor) cavity. Electro-optic techniques, on the other hand, leverage state-of-the-art and mature electronic devices and systems to control the laser frequency precisely. This will push some of the challenges from optical domain into the electrical one where control and manipulation of the signals and systems are comparatively more manageable and cost-effective. \\
A key building block in electro-optic laser stabilization techniques is an optical frequency noise discriminator. An OFND measures frequency fluctuations by comparing to a frequency reference and translates it into an electronic signal that can be processed in the electrical domain. Different OFND configurations have been explored such as "squash" locking technique \cite{diorico2022laser, chabbra2021high}, the PDH laser stabilization method \cite{drever1983laser}, and the unbalanced MZI \cite{sorin1992frequency, idjadi2020nanophotonic}. The PDH loop stands out as the most well-known precision laser instrumentation technique among the extensively utilized OFNDs \cite{liu202236, spencer2016stabilization, wang2020active, kelleher2023compact}. Using the PDH technique, a sharp asymmetric error signal can be generated, which can then be utilized as a servo signal to stabilize the laser frequency. Despite excellent performance, the PDH scheme requires electro-optic phase modulator and relatively fast and complex electronics for modulation and demodulation which increases the power consumption and area for an integrated PDH chip \cite{idjadi2019laser, idjadi2017integrated}. Alternatively, a passive-only unbalanced MZI can serve as an OFND where the two arms of the MZI are phase locked at the quadrature point \cite{bagheri2009semiconductor}. Although the unbalanced MZI has a simple architecture to implement on a chip, achieving high-frequency detection sensitivity comparable to the PDH method requires either a large optical delay line or a substantial electronic gain that comes at the cost of chip area of overall system power consumption.\\ 
Here, we propose and experimentally demonstrate a modulation-free laser stabilization technique using a cavity-coupled MZI on a silicon photonic chip as an OFND. The proposed frequency noise discrimination technique utilizes a high Q-factor cavity coupled to a MZI which breaks the trade-off between sensitivity, complexity, chip area, and power consumption.

With a careful design of the on-chip high Q-factor cavity coupled to an MZI, 4 orders of magnitude suppression in frequency noise of a commercially available distributed feedback (DFB) laser is achieved. On-chip thermal tuners are implemented for the potential trimming of fabrication-induced errors and also for facilitating the broadband operation of the OFND. The proposed architecture combines the advantages of a passive-only structure, which offers simplicity and less electronic power consumption for error signal processing, with the high sensitivity of the widely utilized PDH technique. The proof-of-concept integrated photonic chip is fabricated in AIM Photonic commercially available 180 nm silicon-on-insulator (SOI) process. The photonic chip occupies an area of 0.456 mm$^2$ and consumes only about 50 $\mu$W power for reverse biasing the balanced photo detector. The proposed architecture offers a promising solution for achieving sensitive, simple, and low power laser frequency stabilization systems and sets the stage for the development of low-cost, scalable, and stable integrated lasers.

\section{Results}\label{sec2_result}

\subsection{The principle of operation}
\begin{figure}[b]
    \centering
    \includegraphics[width=\textwidth]{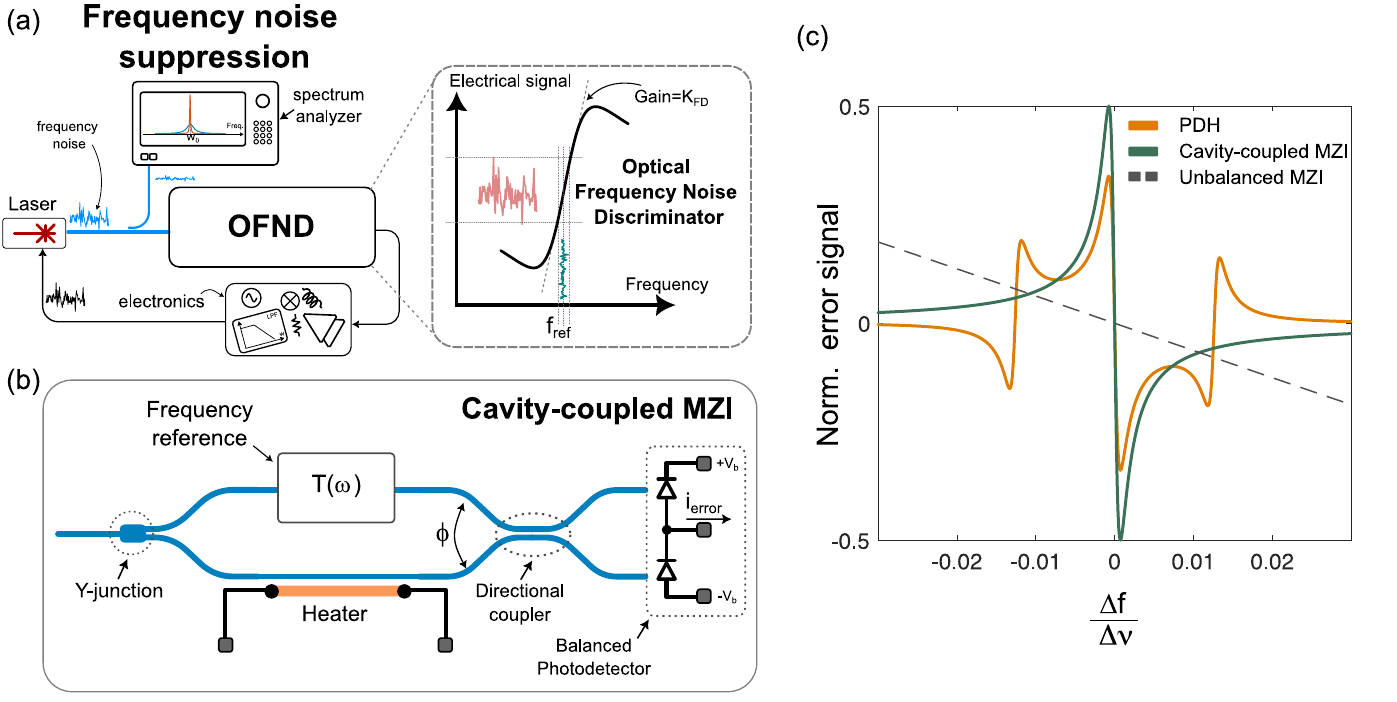}
    \caption{\textbf{The cavity-coupled MZI frequency discriminator.} \small (a) The block diagram of an electro-optic laser frequency stabilization using an optical frequency noise discriminator (OFND). The OFND response is asymmetric around the frequency reference point,$f_{ref}$, and hence small frequency fluctuations translate into electrical signal by the OFND gain. (b) The conceptual diagram of the proposed cavity-coupled MZI. $T(\omega)$ is the transfer function of a frequency reference (\textit{e.g.} optical cavity) coupled into the top arm of the MZI. (c) Numerical analysis comparing the normalized error signal for the proposed cavity-coupled MZI, a conventional unbalanced MZI and the PDH architecture. $\Delta f$ and $\Delta \nu$ are the laser offset frequency compared to $f_{ref}$ and the free-spectral range of the cavity.}
    \label{fig1_intro}
\end{figure}
Figure \ref{fig1_intro}(a) shows the block diagram of an electro-optic laser frequency noise reduction loop. As mentioned earlier, the key part of this loop is an OFND that senses the frequency fluctuations of the incoming laser signal, compares it with an optical frequency reference ($f_{ref}$ ), and generates an electronic signal whose amplitude is proportional to the intensity of frequency fluctuations. This error signal is amplified in the electrical domain and fed back into the laser to stabilize its frequency. The details of the close-loop operation and the linearized block diagram of the control loop is discussed comprehensively in Supplementary Note 1. The OFND can mathematically be represented by an asymmetric transfer function where the small frequency perturbation around $f_{ref}$ is amplified by a gain ($K_{FD}$) and converted into an electronic error signal. The slope of the transfer function also indicates the sensitivity of the OFND in measuring frequency noise.\\
 Figure \ref{fig1_intro}(b) shows the proposed cavity-coupled MZI as an OFND that maintains a simple passive structure without any need for fast optical phase modulation. In this method, the incoming laser intensity is split equally into two MZI branches using a broadband Y-junction. The top branch of the MZI is coupled into an optical frequency reference (\textit{e.g.} an optical resonator or a cavity). The amplitude and phase of the electric field at the output of the cavity is affected by that of the frequency reference, $T(\omega)$. The bottom arm of the MZI is used to interfere with the frequency reference output electric field using a directional coupler. a balanced photodetector is used to photodetect the output of the MZI and by subtracting the currents, the error signal ($i_{error}$) is generated. In other words, the proposed architecture is a coherent detector that uses the input laser signal (bottom arm of the MZI) to down-convert the signal at the output of the cavity. In this way the sharp asymmetric phase transition in the transfer function of the cavity at frequency of $f_{ref}$ will translate in a sharp electrical error signal that can be used to lock the laser in a feedback loop. 
As discussed in Supplementary Note 2, the error signal can be written as
\begin{equation}
    i_{error}(\omega)=RP_0|T(\omega)|\times Sin(\psi(\omega)-\phi), \label{eq1_error}
\end{equation}
where $|T(\omega)|,\psi(\omega), \phi, P_0$, and $R$ are the amplitude and phase of the optical reference at the frequency of $\omega$, phase difference between arms of the MZI controlled by a thermal phase shifter, intensity of the electric field at the input of the MZI, and the responsivity of the photodetectors, respectively. Figure \ref{fig1_intro}(c) is the numerical analysis of the normalized error signal of the PDH, a conventional unbalanced MZI, and the proposed cavity-coupled MZI structure using Eq. (\ref{eq1_error}). To ensure a fair comparison, assuming a fixed area to implement the OFND, a 1 mm circumference ring resonator is used as the frequency reference in both the PDH and the cavity-coupled MZI. The length mismatch between the arms of the unbalanced MZI is also set to 1 mm . The waveguide loss is assumed 0.2 dB/cm. The details of the numerical comparison is presented in Supplementary Note 3. As shown in Fig. \ref{fig1_intro}(c), the error signal of the cavity-coupled MZI is significantly sharper than the conventional unbalanced MZI for a same given setting, 
 and indeed is comparable to that of the PDH offering the same level of sensitivity but much simpler architecture.
\begin{figure*}[tb]
    \centering
    \includegraphics[width=0.7\textwidth]{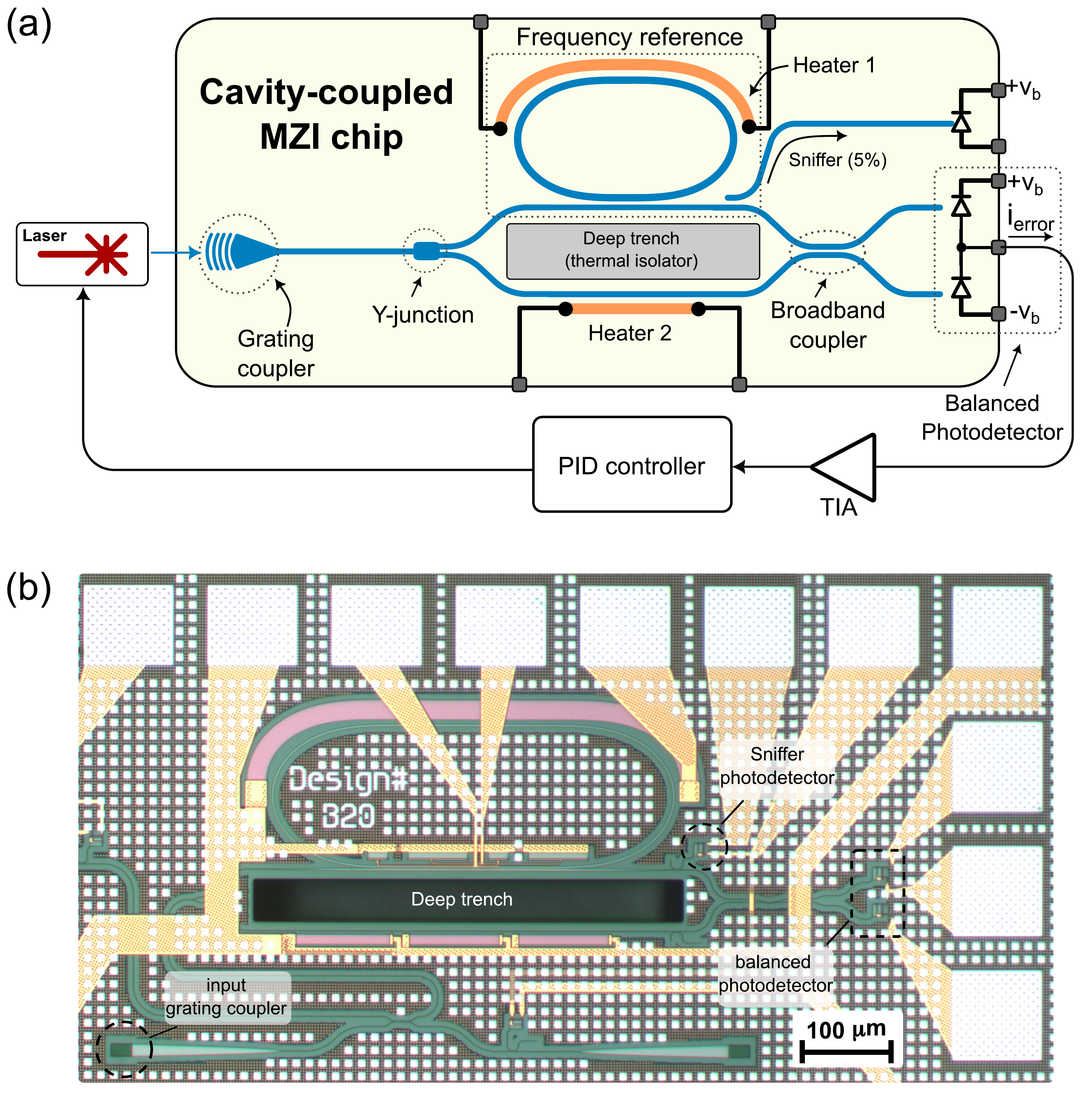}
    \caption{\textbf{The modulation-free laser stabilization scheme.} \small (a) The proposed cavity-coupled MZI is used as an OFND in a feedback loop for laser frequency noise suppression. A high Q-factor Euler microring resonator is utilized as a frequency reference. Half of the on-chip laser intensity is injected to the microring resonator and coherently interfered with the bottom branch and converted to electrical signal using balanced Germanium photodetectors. A small portion of the ring resonator output is used to monitor its response. The generated error signal, $i_{error}$, is then amplified, and fed into the PID controller. (b) The micro-photograph of the integrated photonic chip in AIM Photonics 180 nm silicon-on-insulator (SOI) process. The size of the photonic integrated circuit measures 0.95 mm $\times$ 0.48 mm. PID: proportional-integral-derivative, TIA: trans-impedance amplifier. }
    \label{fig2_concept}
\end{figure*}
\subsection{Modulation-free laser frequency noise suppression system}
Figure \ref{fig2_concept}(a) shows the block diagram of the implemented modulation-free laser stabilization scheme using the implemented cavity-coupled MZI as an OFND. As illustrated in Fig. \ref{fig2_concept}(a), the laser output is coupled into the chip via a grating coupler and divided in half by a Y-junction. In the top branch, a high Q-factor microring resonator is used as an optical frequency reference which filters the amplitude and phase of the incoming light. The phase response of the ring resonator exhibits rapid and asymmetric change around its resonance. When combined with the amplitude response, this characteristic provides sufficient information to determine the offset between the laser frequency and $f_{ref}$, as well as whether the laser frequency is higher or lower than $f_{ref}$.
The output of the MZI uses a 2x2 adiabatic broadband coupler\cite{chen2017broadband} terminated to balanced photodetectors. As described in detail in Supplementary Note 2, the output error signal is asymmetric around $f_{ref}$ and can be used as a servo signal. 
The error signal is amplified and converted to voltage using an off-chip trans-impedance amplifier (TIA). The voltage signal is then fed into a proportional-integral-derivative (PID) controller, which modulates the laser current and corrects any frequency error. To compensate for potential fabrication-induced errors and adjust the microring resonance frequency and the optical phase of the bottom MZI branch, thermal phase shifters that are thermally isolated by a deep trench  are utilized. Figure \ref{fig2_concept}(b) shows the micro-photograph of the photonic integrated circuit implemented in AIM Photonics 180 nm SOI process. The silicon photonic chip area is 0.456 mm$^2$.

\subsection{High Q-factor silicon microring resonator}

\begin{figure*}[t]
    \centering
    \includegraphics[width=0.85\textwidth]{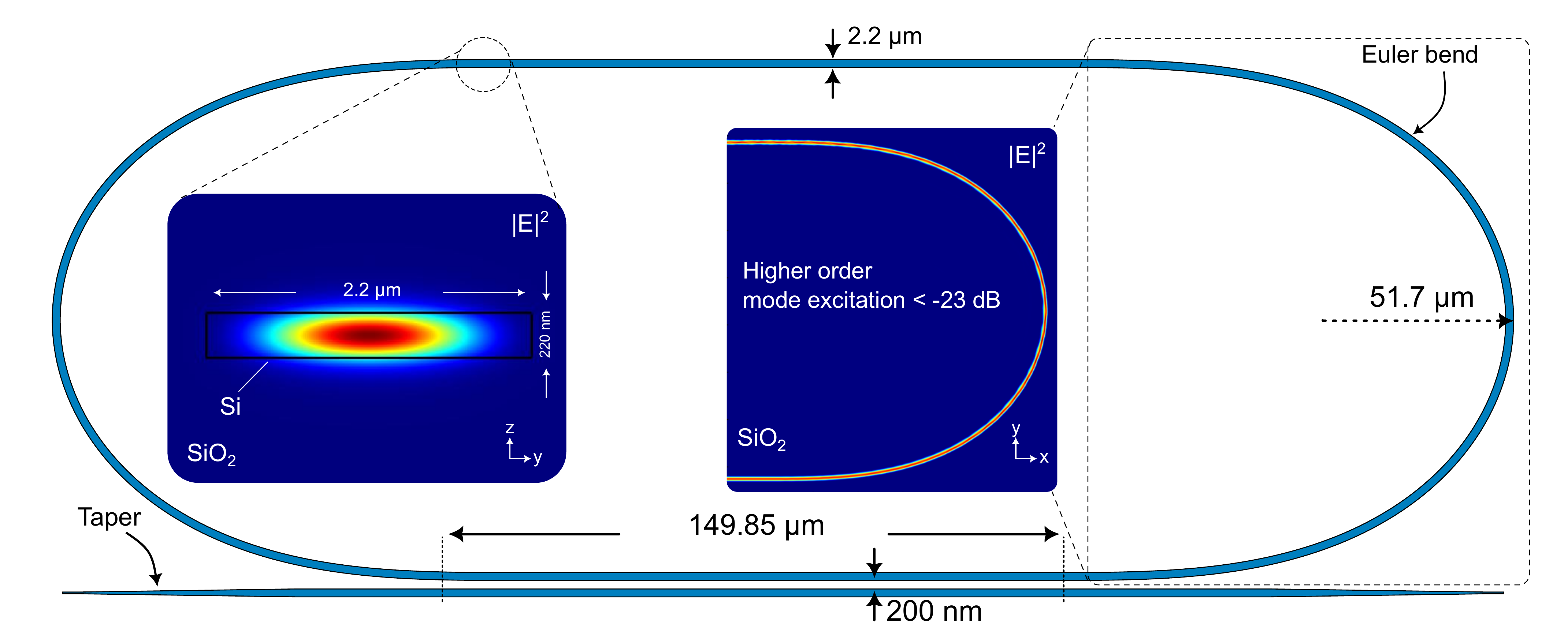}
    \caption{\textbf{High Q-factor silicon microring resonator.} \small In order to minimize the interface scattering loss due to waveguide edge roughness, a wide multi-mode waveguide is utilized. Moreover, to avoid excitation of higher order modes and Euler bend is used. As shown by the Finite-Difference Time-Domain simulation of the waveguide cross-sections (y-z and x-y planes), the fundamental TE-mode is well preserved inside the cavity. The higher order mode excitation is less than -23 dB.}
    \label{fig4_ring}
\end{figure*}

Utilization of a microring resonator as an optical frequency reference plays a crucial role in OFND performance. A stable high Q-factor microring resonator , when coupled to the MZI, effectively enhances the sensitivity of the OFND. This, in turn, directly impacts the closed-loop operation and the ultimate laser frequency noise suppression. It is important to highlight that the proposed architecture can be implemented not only on various material platforms, such as low-loss silicon nitride, but also using bench-top ultra-low expansion and stable etalons. Choosing silicon as a platform to implement the optical frequency reference is a trade-off between different design considerations such as potential co-integration with CMOS electronics, chip area, scalability, cost, and ultimate frequency stability.\\
In order to achieve a high Q-factor microring cavity, it is necessary to minimize intra-cavity losses including silicon nanophotonic waveguides. The propagation loss is influenced by several factors, with interface scattering and bend radiation loss being the most significant in a state-of-the-art silicon photonic foundry process \cite{bauters2011ultra, vlasov2004losses, fahrenkopf2019aim}. The top-bottom surface roughness of a waveguide is well-controlled by the foundry and it is not a design parameter. However, careful design of waveguide width can significantly improve the propagation loss. Figure \ref{fig4_ring} illustrates the implemented microring resonator in silicon. Utilizing multi-mode waveguides can reduce TE-mode overlap with side-wall roughness and greatly enhances the waveguide transmission \cite{bauters2011ultra}. The implemented multi-mode waveguide is 2.2 $\mu$m wide and the theoretical fit model applied to the measured microring response suggests a propagation loss of approximately 0.2 dB/cm. Although wide multi-mode waveguides can greatly reduce interfacial scattering loss, bend radiation loss increases significantly if a tight bend is used \cite{jiang2018low}, especially for a highly multi-mode waveguide. An Euler bend is employed to achieve a compact multi-mode bend with minimal excitation of higher-order modes and mode cross-talk. The ring resonator is designed to achieve critical coupling to maximize OFND sensitivity, however, the fabricated ring are slightly under-coupled due to the potential fabrication-induced errors. Supplementary Note 4 discusses OFND gain sensitivity to waveguide loss and ring coupling ratio.
As illustrated in Fig. \ref{fig4_ring}, the fundamental TE-mode remains preserved within the bent multi-mode waveguide. The implemented microring resonator has a circumference of about 950 $\mu$m which corresponds to a free spectral range (FSR) of about 83.5 GHz with a loaded Q-factor of about 2.5 million at the resonance wavelength of 1550.73 nm.

\subsection{The open-loop operation: device characterization and error signal}
\begin{figure}[bt]
    \centering
    \includegraphics[width=0.6\textwidth]{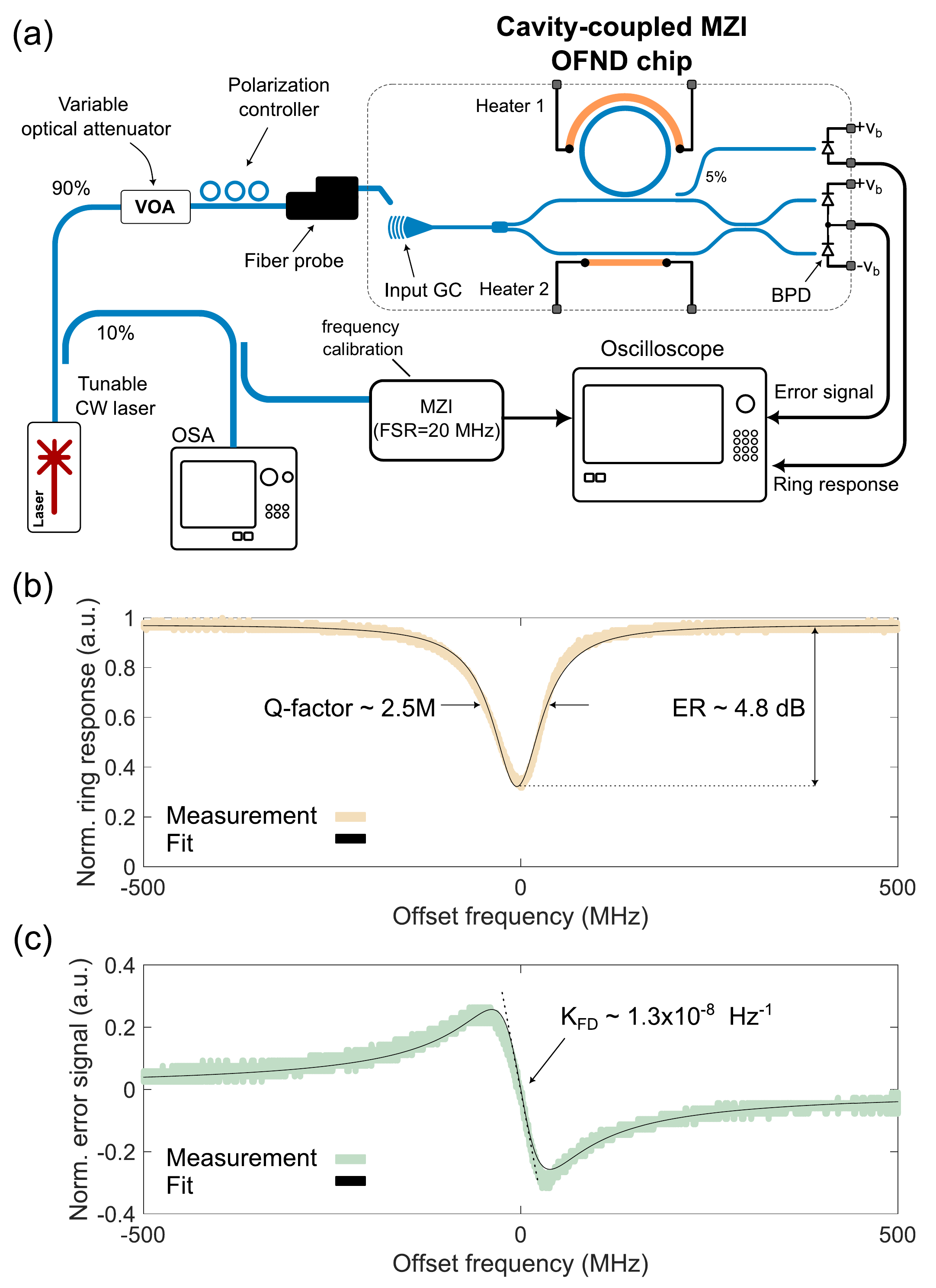}
    \caption{\textbf{The open loop response.} \small (a) Measurement setup for the open loop and microring resonator characterization. (b) The microring resonator response is measured via the top photodetector sniffer. The ring has a measured extinction ratio (ER) of 4.8 dB and a loaded Q-factor of 2.5 million. (c) The measured asymmetric error signal. Heaters are off during the test. CW: continuous wave, OSA: optical spectrum analyzer, GC: grating coupler, BPD: balanced photodetector.}
    \label{fig5_openloop}
\end{figure}
Figure \ref{fig5_openloop}(a) shows the schematic of the measurement setup to characterize the open-loop performance and the error signal. A tunable continues wave laser (TOPTICA CTL 1550) with a wavelength of 1550.7 nm is coupled into the chip via the on-chip grating coupler. To ensure linear operation and thermal stability of the high Q-factor microring resonator, laser power coupled to the grating coupler is set to 0.7 mW using a variable optical attenuator (VOA). The silicon chip temperature is stabilized at 27$^o$C. The laser frequency is continuously scanned within the range of 30 GHz. A calibrated fiber-based MZI with a FSR of 20 MHz is used for time-frequency translation. As shown in Fig. \ref{fig5_openloop}(a), a sniffer photodetector with 5\% coupling ratio is utilized to monitor the resonance response. The resonance response of the ring and the error signal are measured simultaneously using an oscilloscope . Figure \ref{fig5_openloop}(b) shows the response of the ring where the Q-factor is about 2.5 million. The measured extinction ratio is about 4.8 dB that suggests the fabricated microring is slightly under coupled, likely due to fabrication error in the coupling gap. Figure \ref{fig5_openloop}(c) shows the measured normalized asymmetric error signal that suggests OFND sensitivity of 1.3$\times$10$^{-8}$ Hz$^{-1}$ which agrees well with analytical models. Since the ring and MZI are optimized for 1550 nm, there was no need to utilize Heaters 1 and 2 during the measurement. 
\subsection{The closed-loop operation and laser frequency noise suppression}
\begin{figure}[bt]
    \centering
    \includegraphics[width=0.6\textwidth]{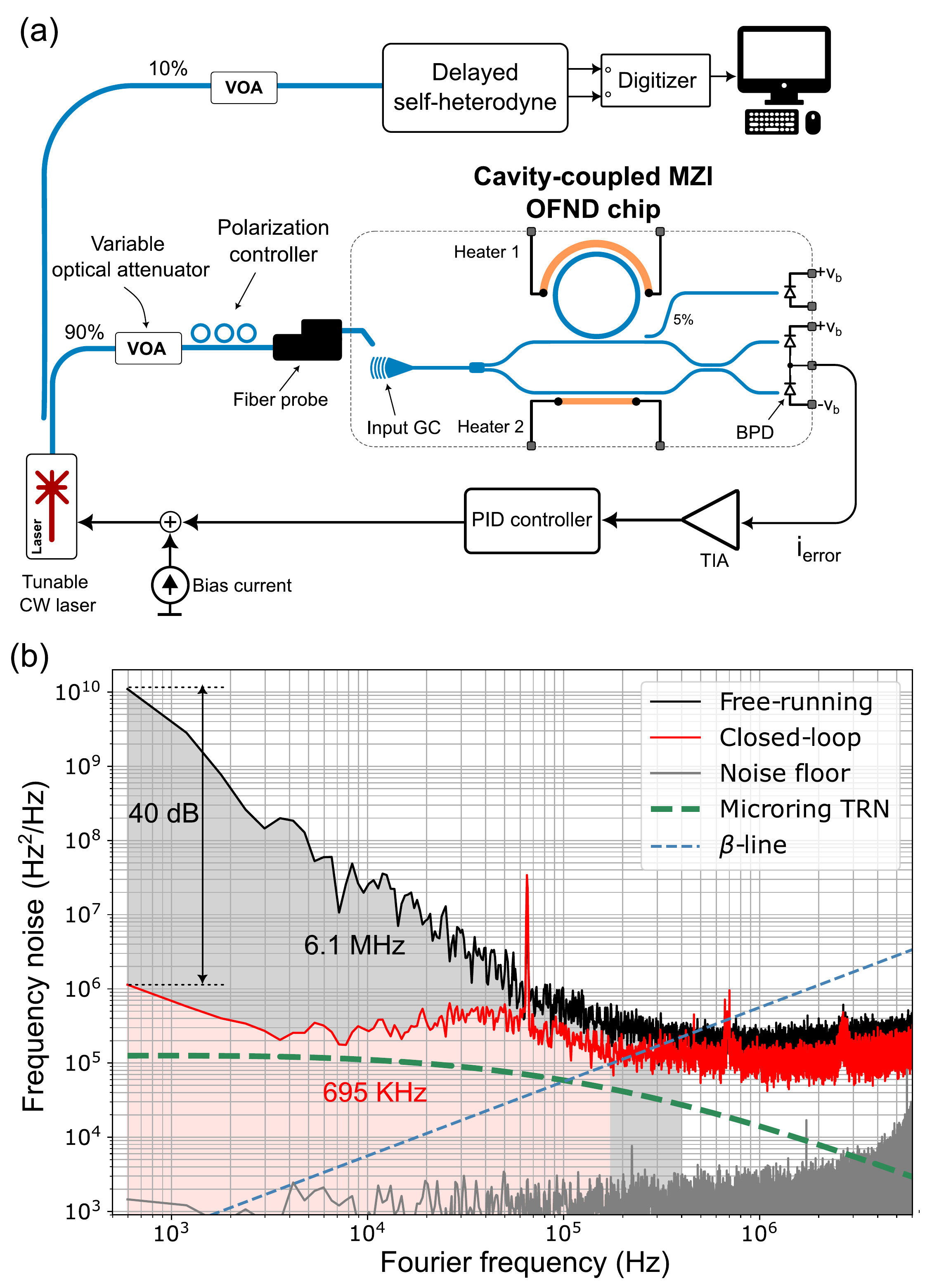}
    \caption{\textbf{The closed-loop operation.} \small (a) Measurement setup for closed-loop operation and laser locking experiment. The small portion of the laser output power is used in the delayed self-heterodyne (DSH) for frequency noise measurement. Heaters are off during this measurement.(b) The power spectral density of frequency noise of the AeroDiode DFB laser under free-running and closed-loop operation. The highlighted regions indicate frequency band which contribute to the laser linewidth. The frequency noise of the DFB laser is suppressed by 40 dB at low Fourier frequencies and is limited to the silicon microring thermorefractive noise (TRN). VOA: variable optical attenuator, TIA: trans-impedance amplifier, CW: continues wave, GC: grating coupler.}
    \label{fig6_closeloop}
\end{figure}
As mentioned previously, the asymmetric error signal can serve as the servo signal to precisely adjust and lock the laser frequency to the resonance frequency of the cavity. Figure \ref{fig6_closeloop} (a) shows the block diagram of the closed-loop measurement setup. As shown in Fig. \ref{fig6_closeloop}(a), a DFB laser (AeroDiode 1550LD-2-0-0-1) with a free-running integral linewidth of 6.1 MHz is used.
The laser output is directed into the chip by utilizing a 90\%/10\% coupler, which is then followed by a VOA. This arrangement allows for the adjustment of the coupled power into the chip to about 0.5 mW. The generated error signal is amplified using a low noise TIA with a gain of 5 K$\Omega$ and total input-referred current noise of 3.4 pA/$\sqrt{Hz}$ which is three orders of magnitude lower than the thermorefractive noise (TRN) of the microring. Part of the laser power is used in a fiber-based delayed self-heterodyne (DSH) interferometer and the output signal is digitized and sent to a computer for processing \cite{yuan2022correlated}. Figure \ref{fig6_closeloop}(b) shows the frequency noise measurement results for the free-running and locked laser. As shown in Fig. \ref{fig6_closeloop}(b), the cavity-coupled MZI OFND can effectively suppress the frequency noise of the free-running DFB laser by 40 dB at low Fourier frequencies. The frequency noise reduction bandwidth is about 300 KHz which is mainly limited by the frequency modulation response of the laser \cite{vankwikelberge1989analysis} and the modulation bandwidth of the current driver ($<$1 MHz). The $\beta$-separation line indicates highlighted regions that contribute to the laser linewidth \cite{di2010simple}. As shown by the highlighted regions in Fig. \ref{fig6_closeloop}(b), the integral linewidth of the free-running laser is reduced from 6.1 MHz to 695 KHz under closed-loop condition. It is worth mentioning that the suppressed frequency noise is limited to the TRN of the silicon microring resonator \cite{huang2019thermorefractive, panuski2020fundamental}. 

\section{Discussion}\label{sec3_conclusion}
Ideally, a cavity that supports a larger mode volume and is made from a temperature insensitive material can enable a lower TRN limit, resulting in a more stable laser. Ultra-low loss silicon nitride and micro-fabricated mirrors \cite{jin2022micro} present themselves as promising technology platforms for fulfilling these requirements. However, this improvement comes at the cost of chip area, packaging complexity, and availability of the technology for large-scale and robust integrated photonic chip manufacturing. 
On the other hand, silicon photonic platforms are evolving rapidly and while silicon may not be the most suitable choice for achieving highly stable micro-cavities, it certainly serves as an excellent alternative in numerous applications where strict level of laser stability may not be required. The combination of robust large-scale silicon photonic manufacturing and the ability to monolithically integrate with mature CMOS electronics provides integrated silicon photonics with a significant advantage over other alternatives. It is important to note that the choice of technology depends on the specific application. Our proposed architecture can be implemented on a suitable technology platform to effectively meet the requirements of an application.

In conclusion, we have experimentally demonstrated the first modulation-free laser frequency noise stabilization using integrated cavity-coupled MZI. The integrated cavity-coupled MZI consists of a high-Q Euler microring resonator with a loaded quality factor of about 2.5 million and extinction ratio of about 4.8 dB which suggests normalized OFND gain of 1.3$\times$10$^{-8}$ Hz$^{-1}$. The cavity-coupled MZI is used to suppress the frequency noise of a commercially available DFB laser by 40 dB at 1 KHz offset frequency. This corresponds to the integral linewidth reduction of the free-running laser from 6.1 MHz to 695 KHz. The implemented chip occupies 0.456 mm$^2$ and is integrated on AIM Photonics 180 nm process. 
\section{Methods}\label{sec4_method}
\subsection{Photonic chip implementation}
All photonic devices are monolithically integrated on AIM Photonics commercially available 180 nm SOI process. The laser is coupled into the chip vertically using a grating coupler. A Y-junction with excess loss of less than 0.5 dB is utilized at the input of the MZI to split the laser intensity equally. A high Q-factor microring resonator is coupled to the top arm of the MZI. The ring resonator is made with wide multi-mode waveguides and Euler bends to both reduce the interfacial scattering loss and avoid excitation of higher-order modes. Thermal phase shifters are integrated into the photonic chip for potential adjustment of phase or compensation for fabrication-induced errors. The two arms of the MZI are combined using an adiabatic broadband directional coupler. A balanced photodetector is used to generate the error signal. The responsivity and dark current of the on-chip Germanium photodetector at 2.5 V reverse bias voltage are 1.16 A/W (at 1550 nm) and 40 nA, respectively. The photonic chip occupies area of 0.456 mm$^2$.
\subsection{The closed-loop operation in presence of noise sources}
Different noise sources contribute to the ultimate frequency noise limit. The two main sources of noise are the TRN of the reference cavity and the total electronic noise that includes input referred noise of electronics and shot noise of balanced photodetector. The closed-loop operation and minimum achievable frequency noise in the presence of these noise sources are discussed in Supplementary Note 1.
\subsection{Delayed self-heterodyne frequency noise measurement}
A fiber-based delayed self-heterodyne setup is built for laser frequency noise measurement. The first null of the MZI is at 8.7 MHz, corresponding to a single-mode fiber length of about 23 m. An acousto-optic frequency shifter is used in the second arm of the MZI to up-convert the laser phase noise to 27 MHz which helps to mitigate electronic flicker noise for better measurement sensitivity. The two arms of the MZI are cross-coupled and fed into two independent low noise TIAs. The outputs of the TIAs are digitized at 250 MS/s and recorded for a duration of 5 msec and sent to a computer for processing.
\subsection{The OFND gain sensitivity analysis}
The design of the high Q-factor microring resonator in the OFND directly impacts the frequency noise detection gain and laser frequency noise suppression. Supplementary Note 4 discusses the sensitivity of OFND gain to both the coupling ratio of the ring and the waveguide propagation loss. \\

\section*{Data availability}
The data that support the plots within this paper and other findings of this study
are available from the corresponding author upon reasonable request.
\section*{Acknowledgements}
We thank Nicolas K. Fontaine and Andrea Blanco-Redondo for helpful discussions and support.
\section*{Author contributions}
M.H.I. conceived the project idea. M.H.I. designed, simulated and laid out the integrated photonic circuits and conducted measurements. M.H.I designed the printed circuit board and the on-board electronics. K.K. packaged the integrated photonic chip. M.H.I. wrote the manuscript. 
\section*{Competing interests}
The authors declare no competing interests.


\begin{thebibliography}{32}
\expandafter\ifx\csname natexlab\endcsname\relax\def\natexlab#1{#1}\fi
\expandafter\ifx\csname bibnamefont\endcsname\relax
  \def\bibnamefont#1{#1}\fi
\expandafter\ifx\csname bibfnamefont\endcsname\relax
  \def\bibfnamefont#1{#1}\fi
\expandafter\ifx\csname citenamefont\endcsname\relax
  \def\citenamefont#1{#1}\fi
\expandafter\ifx\csname url\endcsname\relax
  \def\url#1{\texttt{#1}}\fi
\expandafter\ifx\csname urlprefix\endcsname\relax\def\urlprefix{URL }\fi
\providecommand{\bibinfo}[2]{#2}
\providecommand{\eprint}[2][]{\url{#2}}

\bibitem[{\citenamefont{Al-Taiy et~al.}(2014)\citenamefont{Al-Taiy, Wenzel,
  Preu{\ss}ler, Klinger, and Schneider}}]{al2014ultra}
\bibinfo{author}{\bibfnamefont{H.}~\bibnamefont{Al-Taiy}},
  \bibinfo{author}{\bibfnamefont{N.}~\bibnamefont{Wenzel}},
  \bibinfo{author}{\bibfnamefont{S.}~\bibnamefont{Preu{\ss}ler}},
  \bibinfo{author}{\bibfnamefont{J.}~\bibnamefont{Klinger}}, \bibnamefont{and}
  \bibinfo{author}{\bibfnamefont{T.}~\bibnamefont{Schneider}},
  \bibinfo{journal}{Optics Letters} \textbf{\bibinfo{volume}{39}},
  \bibinfo{pages}{5826} (\bibinfo{year}{2014}).

\bibitem[{\citenamefont{Blumenthal et~al.}(2020)\citenamefont{Blumenthal,
  Ballani, Behunin, Bowers, Costa, Lenoski, Morton, Papp, and
  Rakich}}]{blumenthal2020frequency}
\bibinfo{author}{\bibfnamefont{D.~J.} \bibnamefont{Blumenthal}},
  \bibinfo{author}{\bibfnamefont{H.}~\bibnamefont{Ballani}},
  \bibinfo{author}{\bibfnamefont{R.~O.} \bibnamefont{Behunin}},
  \bibinfo{author}{\bibfnamefont{J.~E.} \bibnamefont{Bowers}},
  \bibinfo{author}{\bibfnamefont{P.}~\bibnamefont{Costa}},
  \bibinfo{author}{\bibfnamefont{D.}~\bibnamefont{Lenoski}},
  \bibinfo{author}{\bibfnamefont{P.~A.} \bibnamefont{Morton}},
  \bibinfo{author}{\bibfnamefont{S.~B.} \bibnamefont{Papp}}, \bibnamefont{and}
  \bibinfo{author}{\bibfnamefont{P.~T.} \bibnamefont{Rakich}},
  \bibinfo{journal}{Journal of Lightwave Technology}
  \textbf{\bibinfo{volume}{38}}, \bibinfo{pages}{3376} (\bibinfo{year}{2020}).

\bibitem[{\citenamefont{Ludlow et~al.}(2015)\citenamefont{Ludlow, Boyd, Ye,
  Peik, and Schmidt}}]{ludlow2015optical}
\bibinfo{author}{\bibfnamefont{A.~D.} \bibnamefont{Ludlow}},
  \bibinfo{author}{\bibfnamefont{M.~M.} \bibnamefont{Boyd}},
  \bibinfo{author}{\bibfnamefont{J.}~\bibnamefont{Ye}},
  \bibinfo{author}{\bibfnamefont{E.}~\bibnamefont{Peik}}, \bibnamefont{and}
  \bibinfo{author}{\bibfnamefont{P.~O.} \bibnamefont{Schmidt}},
  \bibinfo{journal}{Reviews of Modern Physics} \textbf{\bibinfo{volume}{87}},
  \bibinfo{pages}{637} (\bibinfo{year}{2015}).

\bibitem[{\citenamefont{Li et~al.}(2013)\citenamefont{Li, Lee, and
  Vahala}}]{li2013microwave}
\bibinfo{author}{\bibfnamefont{J.}~\bibnamefont{Li}},
  \bibinfo{author}{\bibfnamefont{H.}~\bibnamefont{Lee}}, \bibnamefont{and}
  \bibinfo{author}{\bibfnamefont{K.~J.} \bibnamefont{Vahala}},
  \bibinfo{journal}{Nature communications} \textbf{\bibinfo{volume}{4}},
  \bibinfo{pages}{2097} (\bibinfo{year}{2013}).

\bibitem[{\citenamefont{Marra et~al.}(2018)\citenamefont{Marra, Clivati,
  Luckett, Tampellini, Kronj{\"a}ger, Wright, Mura, Levi, Robinson, Xuereb
  et~al.}}]{marra2018ultrastable}
\bibinfo{author}{\bibfnamefont{G.}~\bibnamefont{Marra}},
  \bibinfo{author}{\bibfnamefont{C.}~\bibnamefont{Clivati}},
  \bibinfo{author}{\bibfnamefont{R.}~\bibnamefont{Luckett}},
  \bibinfo{author}{\bibfnamefont{A.}~\bibnamefont{Tampellini}},
  \bibinfo{author}{\bibfnamefont{J.}~\bibnamefont{Kronj{\"a}ger}},
  \bibinfo{author}{\bibfnamefont{L.}~\bibnamefont{Wright}},
  \bibinfo{author}{\bibfnamefont{A.}~\bibnamefont{Mura}},
  \bibinfo{author}{\bibfnamefont{F.}~\bibnamefont{Levi}},
  \bibinfo{author}{\bibfnamefont{S.}~\bibnamefont{Robinson}},
  \bibinfo{author}{\bibfnamefont{A.}~\bibnamefont{Xuereb}},
  \bibnamefont{et~al.}, \bibinfo{journal}{Science}
  \textbf{\bibinfo{volume}{361}}, \bibinfo{pages}{486} (\bibinfo{year}{2018}).

\bibitem[{\citenamefont{Liang et~al.}(2015)\citenamefont{Liang, Ilchenko,
  Eliyahu, Savchenkov, Matsko, Seidel, and Maleki}}]{liang2015ultralow}
\bibinfo{author}{\bibfnamefont{W.}~\bibnamefont{Liang}},
  \bibinfo{author}{\bibfnamefont{V.}~\bibnamefont{Ilchenko}},
  \bibinfo{author}{\bibfnamefont{D.}~\bibnamefont{Eliyahu}},
  \bibinfo{author}{\bibfnamefont{A.}~\bibnamefont{Savchenkov}},
  \bibinfo{author}{\bibfnamefont{A.}~\bibnamefont{Matsko}},
  \bibinfo{author}{\bibfnamefont{D.}~\bibnamefont{Seidel}}, \bibnamefont{and}
  \bibinfo{author}{\bibfnamefont{L.}~\bibnamefont{Maleki}},
  \bibinfo{journal}{Nature communications} \textbf{\bibinfo{volume}{6}},
  \bibinfo{pages}{7371} (\bibinfo{year}{2015}).

\bibitem[{\citenamefont{Dahmani et~al.}(1987)\citenamefont{Dahmani, Hollberg,
  and Drullinger}}]{dahmani1987frequency}
\bibinfo{author}{\bibfnamefont{B.}~\bibnamefont{Dahmani}},
  \bibinfo{author}{\bibfnamefont{L.}~\bibnamefont{Hollberg}}, \bibnamefont{and}
  \bibinfo{author}{\bibfnamefont{R.}~\bibnamefont{Drullinger}},
  \bibinfo{journal}{Optics letters} \textbf{\bibinfo{volume}{12}},
  \bibinfo{pages}{876} (\bibinfo{year}{1987}).

\bibitem[{\citenamefont{Drever et~al.}(1983)\citenamefont{Drever, Hall,
  Kowalski, Hough, Ford, Munley, and Ward}}]{drever1983laser}
\bibinfo{author}{\bibfnamefont{R.~W.} \bibnamefont{Drever}},
  \bibinfo{author}{\bibfnamefont{J.~L.} \bibnamefont{Hall}},
  \bibinfo{author}{\bibfnamefont{F.~V.} \bibnamefont{Kowalski}},
  \bibinfo{author}{\bibfnamefont{J.}~\bibnamefont{Hough}},
  \bibinfo{author}{\bibfnamefont{G.}~\bibnamefont{Ford}},
  \bibinfo{author}{\bibfnamefont{A.}~\bibnamefont{Munley}}, \bibnamefont{and}
  \bibinfo{author}{\bibfnamefont{H.}~\bibnamefont{Ward}},
  \bibinfo{journal}{Applied Physics B} \textbf{\bibinfo{volume}{31}},
  \bibinfo{pages}{97} (\bibinfo{year}{1983}).

\bibitem[{\citenamefont{Idjadi and Aflatouni}(2020)}]{idjadi2020nanophotonic}
\bibinfo{author}{\bibfnamefont{M.~H.} \bibnamefont{Idjadi}} \bibnamefont{and}
  \bibinfo{author}{\bibfnamefont{F.}~\bibnamefont{Aflatouni}},
  \bibinfo{journal}{Nature Photonics} \textbf{\bibinfo{volume}{14}},
  \bibinfo{pages}{234} (\bibinfo{year}{2020}).

\bibitem[{\citenamefont{Aflatouni and Hashemi}(2012)}]{aflatouni2012wideband}
\bibinfo{author}{\bibfnamefont{F.}~\bibnamefont{Aflatouni}} \bibnamefont{and}
  \bibinfo{author}{\bibfnamefont{H.}~\bibnamefont{Hashemi}},
  \bibinfo{journal}{Optics letters} \textbf{\bibinfo{volume}{37}},
  \bibinfo{pages}{196} (\bibinfo{year}{2012}).

\bibitem[{\citenamefont{Gorodetsky et~al.}(2000)\citenamefont{Gorodetsky,
  Pryamikov, and Ilchenko}}]{gorodetsky2000rayleigh}
\bibinfo{author}{\bibfnamefont{M.~L.} \bibnamefont{Gorodetsky}},
  \bibinfo{author}{\bibfnamefont{A.~D.} \bibnamefont{Pryamikov}},
  \bibnamefont{and} \bibinfo{author}{\bibfnamefont{V.~S.}
  \bibnamefont{Ilchenko}}, \bibinfo{journal}{JOSA B}
  \textbf{\bibinfo{volume}{17}}, \bibinfo{pages}{1051} (\bibinfo{year}{2000}).

\bibitem[{\citenamefont{Diorico et~al.}(2022)\citenamefont{Diorico, Zhutov, and
  Hosten}}]{diorico2022laser}
\bibinfo{author}{\bibfnamefont{F.}~\bibnamefont{Diorico}},
  \bibinfo{author}{\bibfnamefont{A.}~\bibnamefont{Zhutov}}, \bibnamefont{and}
  \bibinfo{author}{\bibfnamefont{O.}~\bibnamefont{Hosten}},
  \bibinfo{journal}{arXiv preprint arXiv:2203.04550}  (\bibinfo{year}{2022}).

\bibitem[{\citenamefont{Chabbra et~al.}(2021)\citenamefont{Chabbra, Wade, Rees,
  Sutton, Stochino, Ward, Shaddock, and McKenzie}}]{chabbra2021high}
\bibinfo{author}{\bibfnamefont{N.}~\bibnamefont{Chabbra}},
  \bibinfo{author}{\bibfnamefont{A.~R.} \bibnamefont{Wade}},
  \bibinfo{author}{\bibfnamefont{E.~R.} \bibnamefont{Rees}},
  \bibinfo{author}{\bibfnamefont{A.~J.} \bibnamefont{Sutton}},
  \bibinfo{author}{\bibfnamefont{A.}~\bibnamefont{Stochino}},
  \bibinfo{author}{\bibfnamefont{R.~L.} \bibnamefont{Ward}},
  \bibinfo{author}{\bibfnamefont{D.~A.} \bibnamefont{Shaddock}},
  \bibnamefont{and} \bibinfo{author}{\bibfnamefont{K.}~\bibnamefont{McKenzie}},
  \bibinfo{journal}{Optics Letters} \textbf{\bibinfo{volume}{46}},
  \bibinfo{pages}{3199} (\bibinfo{year}{2021}).

\bibitem[{\citenamefont{Sorin et~al.}(1992)\citenamefont{Sorin, Chang, Conrad,
  and Hernday}}]{sorin1992frequency}
\bibinfo{author}{\bibfnamefont{W.~V.} \bibnamefont{Sorin}},
  \bibinfo{author}{\bibfnamefont{K.~W.} \bibnamefont{Chang}},
  \bibinfo{author}{\bibfnamefont{G.~A.} \bibnamefont{Conrad}},
  \bibnamefont{and} \bibinfo{author}{\bibfnamefont{P.~R.}
  \bibnamefont{Hernday}}, \bibinfo{journal}{Journal of Lightwave Technology}
  \textbf{\bibinfo{volume}{10}}, \bibinfo{pages}{787} (\bibinfo{year}{1992}).

\bibitem[{\citenamefont{Liu et~al.}(2022)\citenamefont{Liu, Chauhan, Wang,
  Isichenko, Brodnik, Morton, Behunin, Papp, and Blumenthal}}]{liu202236}
\bibinfo{author}{\bibfnamefont{K.}~\bibnamefont{Liu}},
  \bibinfo{author}{\bibfnamefont{N.}~\bibnamefont{Chauhan}},
  \bibinfo{author}{\bibfnamefont{J.}~\bibnamefont{Wang}},
  \bibinfo{author}{\bibfnamefont{A.}~\bibnamefont{Isichenko}},
  \bibinfo{author}{\bibfnamefont{G.~M.} \bibnamefont{Brodnik}},
  \bibinfo{author}{\bibfnamefont{P.~A.} \bibnamefont{Morton}},
  \bibinfo{author}{\bibfnamefont{R.~O.} \bibnamefont{Behunin}},
  \bibinfo{author}{\bibfnamefont{S.~B.} \bibnamefont{Papp}}, \bibnamefont{and}
  \bibinfo{author}{\bibfnamefont{D.~J.} \bibnamefont{Blumenthal}},
  \bibinfo{journal}{Optica} \textbf{\bibinfo{volume}{9}}, \bibinfo{pages}{770}
  (\bibinfo{year}{2022}).

\bibitem[{\citenamefont{Spencer et~al.}(2016)\citenamefont{Spencer, Davenport,
  Komljenovic, Srinivasan, and Bowers}}]{spencer2016stabilization}
\bibinfo{author}{\bibfnamefont{D.~T.} \bibnamefont{Spencer}},
  \bibinfo{author}{\bibfnamefont{M.~L.} \bibnamefont{Davenport}},
  \bibinfo{author}{\bibfnamefont{T.}~\bibnamefont{Komljenovic}},
  \bibinfo{author}{\bibfnamefont{S.}~\bibnamefont{Srinivasan}},
  \bibnamefont{and} \bibinfo{author}{\bibfnamefont{J.~E.}
  \bibnamefont{Bowers}}, \bibinfo{journal}{Optics express}
  \textbf{\bibinfo{volume}{24}}, \bibinfo{pages}{13511} (\bibinfo{year}{2016}).

\bibitem[{\citenamefont{Wang et~al.}(2020)\citenamefont{Wang, Wei, Li, Kan, and
  Ren}}]{wang2020active}
\bibinfo{author}{\bibfnamefont{Z.}~\bibnamefont{Wang}},
  \bibinfo{author}{\bibfnamefont{H.}~\bibnamefont{Wei}},
  \bibinfo{author}{\bibfnamefont{Y.}~\bibnamefont{Li}},
  \bibinfo{author}{\bibfnamefont{R.}~\bibnamefont{Kan}}, \bibnamefont{and}
  \bibinfo{author}{\bibfnamefont{W.}~\bibnamefont{Ren}},
  \bibinfo{journal}{Optics Letters} \textbf{\bibinfo{volume}{45}},
  \bibinfo{pages}{1148} (\bibinfo{year}{2020}).

\bibitem[{\citenamefont{Kelleher et~al.}(2023)\citenamefont{Kelleher, McLemore,
  Lee, Davila-Rodriguez, Diddams, and Quinlan}}]{kelleher2023compact}
\bibinfo{author}{\bibfnamefont{M.~L.} \bibnamefont{Kelleher}},
  \bibinfo{author}{\bibfnamefont{C.~A.} \bibnamefont{McLemore}},
  \bibinfo{author}{\bibfnamefont{D.}~\bibnamefont{Lee}},
  \bibinfo{author}{\bibfnamefont{J.}~\bibnamefont{Davila-Rodriguez}},
  \bibinfo{author}{\bibfnamefont{S.~A.} \bibnamefont{Diddams}},
  \bibnamefont{and} \bibinfo{author}{\bibfnamefont{F.}~\bibnamefont{Quinlan}},
  \bibinfo{journal}{Optics Express} \textbf{\bibinfo{volume}{31}},
  \bibinfo{pages}{11954} (\bibinfo{year}{2023}).

\bibitem[{\citenamefont{Idjadi and Aflatouni}(2019)}]{idjadi2019laser}
\bibinfo{author}{\bibfnamefont{M.~H.} \bibnamefont{Idjadi}} \bibnamefont{and}
  \bibinfo{author}{\bibfnamefont{F.}~\bibnamefont{Aflatouni}}, in
  \emph{\bibinfo{booktitle}{2019 IEEE Radio Frequency Integrated Circuits
  Symposium (RFIC)}} (\bibinfo{organization}{IEEE}, \bibinfo{year}{2019}), pp.
  \bibinfo{pages}{319--322}.

\bibitem[{\citenamefont{Idjadi and Aflatouni}(2017)}]{idjadi2017integrated}
\bibinfo{author}{\bibfnamefont{M.~H.} \bibnamefont{Idjadi}} \bibnamefont{and}
  \bibinfo{author}{\bibfnamefont{F.}~\bibnamefont{Aflatouni}},
  \bibinfo{journal}{Nature communications} \textbf{\bibinfo{volume}{8}},
  \bibinfo{pages}{1209} (\bibinfo{year}{2017}).

\bibitem[{\citenamefont{Bagheri et~al.}(2009)\citenamefont{Bagheri, Aflatouni,
  Imani, Goel, and Hashemi}}]{bagheri2009semiconductor}
\bibinfo{author}{\bibfnamefont{M.}~\bibnamefont{Bagheri}},
  \bibinfo{author}{\bibfnamefont{F.}~\bibnamefont{Aflatouni}},
  \bibinfo{author}{\bibfnamefont{A.}~\bibnamefont{Imani}},
  \bibinfo{author}{\bibfnamefont{A.}~\bibnamefont{Goel}}, \bibnamefont{and}
  \bibinfo{author}{\bibfnamefont{H.}~\bibnamefont{Hashemi}},
  \bibinfo{journal}{Optics letters} \textbf{\bibinfo{volume}{34}},
  \bibinfo{pages}{2979} (\bibinfo{year}{2009}).

\bibitem[{\citenamefont{Chen et~al.}(2017)\citenamefont{Chen, Ong, Ang, Lim,
  Png, and Tan}}]{chen2017broadband}
\bibinfo{author}{\bibfnamefont{G.~F.} \bibnamefont{Chen}},
  \bibinfo{author}{\bibfnamefont{J.~R.} \bibnamefont{Ong}},
  \bibinfo{author}{\bibfnamefont{T.~Y.} \bibnamefont{Ang}},
  \bibinfo{author}{\bibfnamefont{S.~T.} \bibnamefont{Lim}},
  \bibinfo{author}{\bibfnamefont{C.~E.} \bibnamefont{Png}}, \bibnamefont{and}
  \bibinfo{author}{\bibfnamefont{D.~T.} \bibnamefont{Tan}},
  \bibinfo{journal}{Scientific reports} \textbf{\bibinfo{volume}{7}},
  \bibinfo{pages}{7246} (\bibinfo{year}{2017}).

\bibitem[{\citenamefont{Bauters et~al.}(2011)\citenamefont{Bauters, Heck, John,
  Dai, Tien, Barton, Leinse, Heideman, Blumenthal, and
  Bowers}}]{bauters2011ultra}
\bibinfo{author}{\bibfnamefont{J.~F.} \bibnamefont{Bauters}},
  \bibinfo{author}{\bibfnamefont{M.~J.} \bibnamefont{Heck}},
  \bibinfo{author}{\bibfnamefont{D.}~\bibnamefont{John}},
  \bibinfo{author}{\bibfnamefont{D.}~\bibnamefont{Dai}},
  \bibinfo{author}{\bibfnamefont{M.-C.} \bibnamefont{Tien}},
  \bibinfo{author}{\bibfnamefont{J.~S.} \bibnamefont{Barton}},
  \bibinfo{author}{\bibfnamefont{A.}~\bibnamefont{Leinse}},
  \bibinfo{author}{\bibfnamefont{R.~G.} \bibnamefont{Heideman}},
  \bibinfo{author}{\bibfnamefont{D.~J.} \bibnamefont{Blumenthal}},
  \bibnamefont{and} \bibinfo{author}{\bibfnamefont{J.~E.}
  \bibnamefont{Bowers}}, \bibinfo{journal}{Optics express}
  \textbf{\bibinfo{volume}{19}}, \bibinfo{pages}{3163} (\bibinfo{year}{2011}).

\bibitem[{\citenamefont{Vlasov and McNab}(2004)}]{vlasov2004losses}
\bibinfo{author}{\bibfnamefont{Y.~A.} \bibnamefont{Vlasov}} \bibnamefont{and}
  \bibinfo{author}{\bibfnamefont{S.~J.} \bibnamefont{McNab}},
  \bibinfo{journal}{Optics express} \textbf{\bibinfo{volume}{12}},
  \bibinfo{pages}{1622} (\bibinfo{year}{2004}).

\bibitem[{\citenamefont{Fahrenkopf et~al.}(2019)\citenamefont{Fahrenkopf,
  McDonough, Leake, Su, Timurdogan, and Coolbaugh}}]{fahrenkopf2019aim}
\bibinfo{author}{\bibfnamefont{N.~M.} \bibnamefont{Fahrenkopf}},
  \bibinfo{author}{\bibfnamefont{C.}~\bibnamefont{McDonough}},
  \bibinfo{author}{\bibfnamefont{G.~L.} \bibnamefont{Leake}},
  \bibinfo{author}{\bibfnamefont{Z.}~\bibnamefont{Su}},
  \bibinfo{author}{\bibfnamefont{E.}~\bibnamefont{Timurdogan}},
  \bibnamefont{and} \bibinfo{author}{\bibfnamefont{D.~D.}
  \bibnamefont{Coolbaugh}}, \bibinfo{journal}{IEEE Journal of Selected Topics
  in Quantum Electronics} \textbf{\bibinfo{volume}{25}}, \bibinfo{pages}{1}
  (\bibinfo{year}{2019}).

\bibitem[{\citenamefont{Jiang et~al.}(2018)\citenamefont{Jiang, Wu, and
  Dai}}]{jiang2018low}
\bibinfo{author}{\bibfnamefont{X.}~\bibnamefont{Jiang}},
  \bibinfo{author}{\bibfnamefont{H.}~\bibnamefont{Wu}}, \bibnamefont{and}
  \bibinfo{author}{\bibfnamefont{D.}~\bibnamefont{Dai}},
  \bibinfo{journal}{Optics express} \textbf{\bibinfo{volume}{26}},
  \bibinfo{pages}{17680} (\bibinfo{year}{2018}).

\bibitem[{\citenamefont{Yuan et~al.}(2022)\citenamefont{Yuan, Wang, Liu, Li,
  Shen, Gao, Chang, Jin, Feshali, Paniccia et~al.}}]{yuan2022correlated}
\bibinfo{author}{\bibfnamefont{Z.}~\bibnamefont{Yuan}},
  \bibinfo{author}{\bibfnamefont{H.}~\bibnamefont{Wang}},
  \bibinfo{author}{\bibfnamefont{P.}~\bibnamefont{Liu}},
  \bibinfo{author}{\bibfnamefont{B.}~\bibnamefont{Li}},
  \bibinfo{author}{\bibfnamefont{B.}~\bibnamefont{Shen}},
  \bibinfo{author}{\bibfnamefont{M.}~\bibnamefont{Gao}},
  \bibinfo{author}{\bibfnamefont{L.}~\bibnamefont{Chang}},
  \bibinfo{author}{\bibfnamefont{W.}~\bibnamefont{Jin}},
  \bibinfo{author}{\bibfnamefont{A.}~\bibnamefont{Feshali}},
  \bibinfo{author}{\bibfnamefont{M.}~\bibnamefont{Paniccia}},
  \bibnamefont{et~al.}, \bibinfo{journal}{Optics Express}
  \textbf{\bibinfo{volume}{30}}, \bibinfo{pages}{25147} (\bibinfo{year}{2022}).

\bibitem[{\citenamefont{Vankwikelberge
  et~al.}(1989)\citenamefont{Vankwikelberge, Buytaert, Franchois, Baets,
  Kuindersma, and Fredriksz}}]{vankwikelberge1989analysis}
\bibinfo{author}{\bibfnamefont{P.}~\bibnamefont{Vankwikelberge}},
  \bibinfo{author}{\bibfnamefont{F.}~\bibnamefont{Buytaert}},
  \bibinfo{author}{\bibfnamefont{A.}~\bibnamefont{Franchois}},
  \bibinfo{author}{\bibfnamefont{R.}~\bibnamefont{Baets}},
  \bibinfo{author}{\bibfnamefont{P.}~\bibnamefont{Kuindersma}},
  \bibnamefont{and}
  \bibinfo{author}{\bibfnamefont{C.}~\bibnamefont{Fredriksz}},
  \bibinfo{journal}{IEEE journal of quantum electronics}
  \textbf{\bibinfo{volume}{25}}, \bibinfo{pages}{2239} (\bibinfo{year}{1989}).

\bibitem[{\citenamefont{Di~Domenico et~al.}(2010)\citenamefont{Di~Domenico,
  Schilt, and Thomann}}]{di2010simple}
\bibinfo{author}{\bibfnamefont{G.}~\bibnamefont{Di~Domenico}},
  \bibinfo{author}{\bibfnamefont{S.}~\bibnamefont{Schilt}}, \bibnamefont{and}
  \bibinfo{author}{\bibfnamefont{P.}~\bibnamefont{Thomann}},
  \bibinfo{journal}{Applied optics} \textbf{\bibinfo{volume}{49}},
  \bibinfo{pages}{4801} (\bibinfo{year}{2010}).

\bibitem[{\citenamefont{Huang et~al.}(2019)\citenamefont{Huang, Lucas, Liu,
  Raja, Lihachev, Gorodetsky, Engelsen, and
  Kippenberg}}]{huang2019thermorefractive}
\bibinfo{author}{\bibfnamefont{G.}~\bibnamefont{Huang}},
  \bibinfo{author}{\bibfnamefont{E.}~\bibnamefont{Lucas}},
  \bibinfo{author}{\bibfnamefont{J.}~\bibnamefont{Liu}},
  \bibinfo{author}{\bibfnamefont{A.~S.} \bibnamefont{Raja}},
  \bibinfo{author}{\bibfnamefont{G.}~\bibnamefont{Lihachev}},
  \bibinfo{author}{\bibfnamefont{M.~L.} \bibnamefont{Gorodetsky}},
  \bibinfo{author}{\bibfnamefont{N.~J.} \bibnamefont{Engelsen}},
  \bibnamefont{and} \bibinfo{author}{\bibfnamefont{T.~J.}
  \bibnamefont{Kippenberg}}, \bibinfo{journal}{Physical Review A}
  \textbf{\bibinfo{volume}{99}}, \bibinfo{pages}{061801}
  (\bibinfo{year}{2019}).

\bibitem[{\citenamefont{Panuski et~al.}(2020)\citenamefont{Panuski, Englund,
  and Hamerly}}]{panuski2020fundamental}
\bibinfo{author}{\bibfnamefont{C.}~\bibnamefont{Panuski}},
  \bibinfo{author}{\bibfnamefont{D.}~\bibnamefont{Englund}}, \bibnamefont{and}
  \bibinfo{author}{\bibfnamefont{R.}~\bibnamefont{Hamerly}},
  \bibinfo{journal}{Physical Review X} \textbf{\bibinfo{volume}{10}},
  \bibinfo{pages}{041046} (\bibinfo{year}{2020}).

\bibitem[{\citenamefont{Jin et~al.}(2022)\citenamefont{Jin, McLemore, Mason,
  Hendrie, Luo, Kelleher, Kharel, Quinlan, Diddams, and Rakich}}]{jin2022micro}
\bibinfo{author}{\bibfnamefont{N.}~\bibnamefont{Jin}},
  \bibinfo{author}{\bibfnamefont{C.~A.} \bibnamefont{McLemore}},
  \bibinfo{author}{\bibfnamefont{D.}~\bibnamefont{Mason}},
  \bibinfo{author}{\bibfnamefont{J.~P.} \bibnamefont{Hendrie}},
  \bibinfo{author}{\bibfnamefont{Y.}~\bibnamefont{Luo}},
  \bibinfo{author}{\bibfnamefont{M.~L.} \bibnamefont{Kelleher}},
  \bibinfo{author}{\bibfnamefont{P.}~\bibnamefont{Kharel}},
  \bibinfo{author}{\bibfnamefont{F.}~\bibnamefont{Quinlan}},
  \bibinfo{author}{\bibfnamefont{S.~A.} \bibnamefont{Diddams}},
  \bibnamefont{and} \bibinfo{author}{\bibfnamefont{P.~T.}
  \bibnamefont{Rakich}}, \bibinfo{journal}{Optica}
  \textbf{\bibinfo{volume}{9}}, \bibinfo{pages}{965} (\bibinfo{year}{2022}).

\end{thebibliography}

\end{document}


\title{Supplementary Information: \\ Modulation-free Laser Stabilization Technique Using Integrated Cavity-Coupled Mach-Zehnder Interferometer}
\author{Mohamad Hossein Idjadi\textsuperscript{*}}
\author{Kwangwoong Kim}

\affiliation{Nokia Bell Labs, 600 Mountain Ave, Murray Hill, NJ 07974, USA.}

\author{\footnotesize\textsuperscript{*}Corresponding author: mohamad.idjadi@nokia-bell-labs.com}

\pacs{}

\maketitle

\section*{Supplementary Note 1: the closed-loop operation in presence of noise sources}
\begin{figure}[b]
    \centering
    \includegraphics[width=0.6\textwidth]{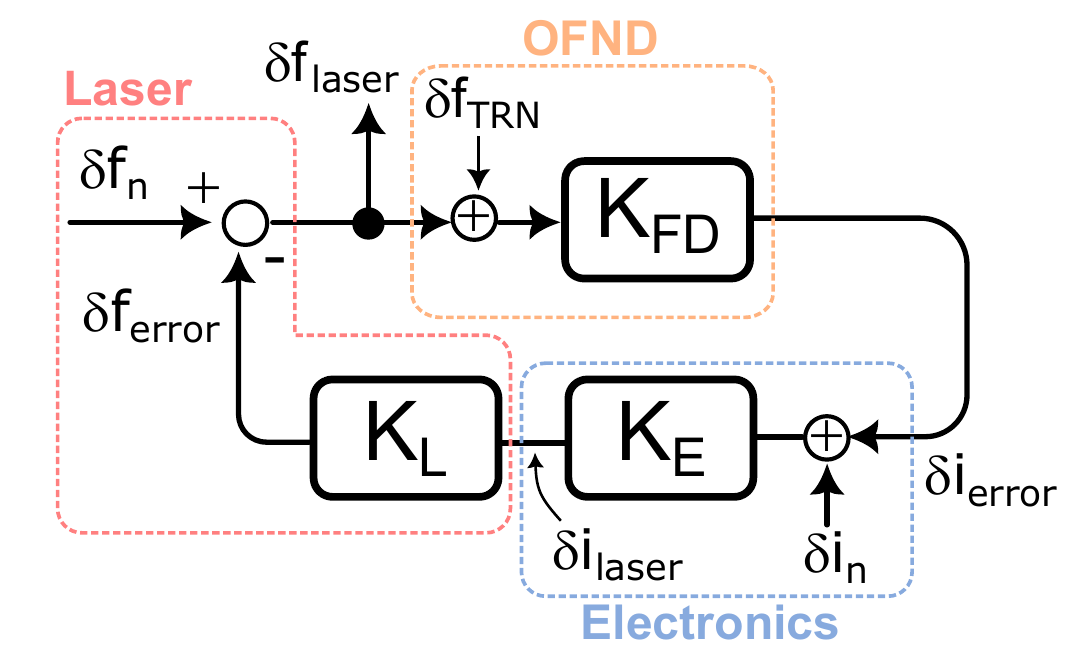}
    \caption{\textbf{The linearized closed-loop block diagram.} $\delta i_n, \delta f_n$, and $\delta f_{TRN}$ are input refereed noise of electronics, intrinsic frequency noise of laser, and thermorefractive noise of the cavity used in the OFND.}
    \label{supp_fig_linearblock}
\end{figure}

Supplementary Figure \ref{supp_fig_linearblock} shows the simplified block diagram of the closed-loop laser frequency noise suppression. The control loop is linearized around the reference frequency ($f_{ref}$), in presence of different noise sources. As shown in Supplementary Figure \ref{supp_fig_linearblock}, there are three main noise sources in the loop contributing to the closed-loop noise performance. These include laser intrinsic frequency noise ($\delta f_n$), thermorefractive noise (TRN, $\delta f_{TRN}$) of the frequency reference used in the optical frequency noise discriminator (OFND), and the total electronics noise ($\delta i_n$) that includes input referred noise of electronics and shot noise of balanced photodetector. The small signal current perturbation due to the frequency fluctuations is
\begin{equation}
    \delta i_{error}=K_{FD}\times (\delta f_{laser} + \delta f_{TRN}), \label{supp_eq1}
\end{equation}
where $\delta f_{laser}$ and $K_{FD}$ [Hz/A] are laser frequency noise under closed-loop condition and conversion gain of OFND, respectively.
The small change in the error signal is amplified by electronic gain, $K_E$ [A/A]. The total current noise at the input of the electronics, $\delta i_n$, includes the input-referred current noise of the amplifiers and shot noise of the balanced photodetector. The laser current perturbation is
\begin{equation}
    \delta i_{laser}=K_E \times (\delta i_n + \delta i_{error}). \label{supp_eq2}
\end{equation}
The laser current perturbations modulate the frequency of the laser by current to frequency conversion gain $K_L$ [Hz/A], 
\begin{equation}
    \delta f_{error}=K_L \times \delta i_{laser}, \label{supp_eq3}
\end{equation}
where $\delta f_{error}$ is the small perturbation in the frequency of the laser.
Under the closed-loop operation, 
\begin{equation}
    \delta f_{laser} = \delta f_n - \delta f_{error}. \label{supp_eq4}
\end{equation}
Using Supplementary Equations (\ref{supp_eq1}) to (\ref{supp_eq4}) (assuming $K_E K_L K_{FD}>>1$), 
\begin{equation}
    \delta f_{laser} \approx (\frac{1}{K_E K_L K_{FD}})\delta f_n-(\frac{1}{K_{FD}})\delta i_n - \delta f_{TRN}. \label{supp_eq5}
\end{equation}
Since $\delta f_n, \delta i_n$, and $\delta f_{TRN}$ are independent random variables with an average of zero, the power spectral density (PSD) of the laser frequency noise is
\begin{equation}
    S_{laser}(f) \approx (\frac{1}{K_E K_L K_{FD}})^2 S_0(f)+(\frac{1}{K_{FD}})^2 S_n(f) + S_{TRN}(f), \label{supp_eq6}
\end{equation}
where $S_{laser}(f), S_0(f), S_n(f)$, and $S_{TRN}(f)$ are the power spectral densities of stabilized laser frequency noise, free-running laser frequency noise, electronic noise, and cavity TRN, respectively. As suggested by Supplementary Equation (\ref{supp_eq6}), the PSD of the free-running laser is suppressed by the open-loop gain. It is worth noting that the modulation gain of laser, $K_L$ can not be changed after laser fabrication. Although the open loop gain can be increased electronically, it comes at the cost of loop bandwidth, power consumption, and even loop stability. Utilizing a sensitive OFND can significantly enhance frequency noise detection gain ($K_{FD}$) and enhance the frequency noise suppression. Moreover, the ultimate laser frequency noise under closed-loop operation is limited by 
\begin{equation}
    S_{limit}(f)=(\frac{1}{K_{FD}})^2 S_n(f) + S_{TRN}(f). \label{supp_eq7}
\end{equation}
According to Supplementary Equation (\ref{supp_eq7}), the TRN of the cavity can impose a limit on the achievable stability. Therefore, considering the specific requirements of an application, the ultimate achievable frequency noise can be engineered by a careful design and proper selection of the material platform to implement the frequency reference \cite{huang2019thermorefractive}.
\newpage
\section*{Supplementary Note 2: the enhanced frequency noise discriminator using cavity-coupled MZI}\label{supp_note_error_signal}

\begin{figure}[b]
    \centering
    \includegraphics[width=\textwidth]{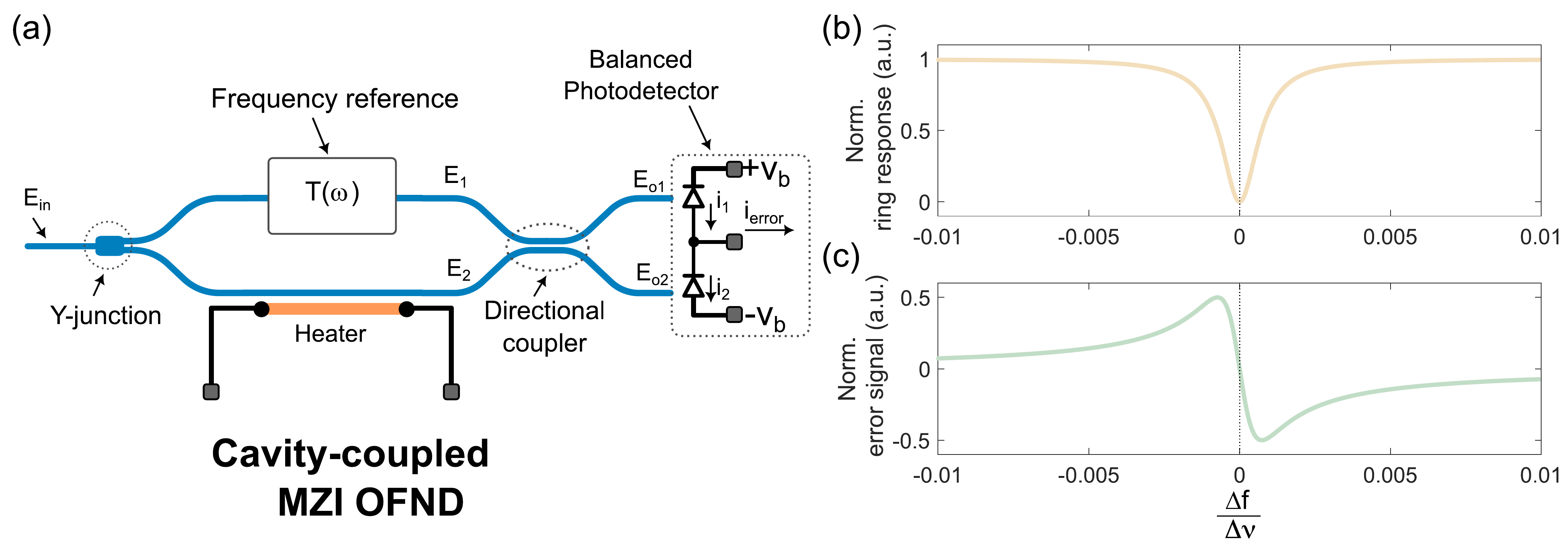}
\caption{\textbf{The cavity-coupled MZI OFND.} (a) The block diagram of the proposed cavity-coupled MZI optical frequency noise discriminator (OFND) architecture. (b) The simulated ring resonator response. (c) The asymmetric error signal. The normalized gain of the OFND is 1.7$\times$10$^{-8}$ Hz$^{-1}$. $\Delta f$ and $\Delta \nu$ are laser frequency offset compared to $f_{ref}$ and the free-spectral range (FSR) of the ring resonator, respectively.}
    \label{supp_fig_simulation_error}
\end{figure}

Supplementary Figure \ref{supp_fig_simulation_error}(a) shows the block diagram of a generalized cavity-coupled MZI OFND architecture. Within this model, we consider a generalized transfer function of an optical frequency reference, $T(.)$. The electric field at the input of the OFND can be expressed as  
\begin{equation}
    E_{in}(t)=\sqrt{P_0} e^{jwt}, \label{supp_eq8}
\end{equation}
where $\omega$ and $P_0$ are the instantaneous laser frequency and laser power, respectively. A Y-junction is used to split the power equally. The electric fields before the output coupler are
\begin{gather}
    E_1(t)=\sqrt{\frac{P_0}{2}} T(\omega)e^{jwt}, \label{supp_eq9} \\
    E_2(t)=\sqrt{\frac{P_0}{2}} e^{j\phi}e^{jwt}, \label{supp_eq10}
\end{gather}
where $T(\omega)$ and $\phi$ are the complex optical reference transfer function and the phase shift introduced by the thermal phase shifter at the bottom branch of MZI, respectively. A directional coupler combines the electric field. The interfered output fields at the top and bottom ports can be written as
\begin{gather}
    E_{o1}(t)=\frac{1}{\sqrt{2}}(E_1(t)+jE_2(t)), \label{supp_eq11}\\
    E_{o2}(t)=\frac{1}{\sqrt{2}}(E_2(t)+jE_1(t)). \label{supp_eq12}
\end{gather}
The optical signal at the output of the MZI is photodetected using a balanced photodetector. The photocurrents $i_1$ and $i_2$ are calculated as
\begin{gather}
    i_1(t)=R|E_{o1}(t)|^2 , \label{supp_eq13}\\
    i_2(t)=R|E_{o2}(t)|^2 , \label{supp_eq14}
\end{gather}
where $R$ is the responsivity of the photodetectors. Using Supplementary Equations (\ref{supp_eq11}) to (\ref{supp_eq14}), the error signal, $i_1(t)-i_2(t)$, can be written as
\begin{equation}
    i_{error}(t)=2R\Im\biggl(E_1(t)E_2(t)^*\biggl), \label{supp_eq15}
\end{equation}
where $\Im(.)$ denotes the imaginary operator. Combining Supplementary Equations (\ref{supp_eq9}),(\ref{supp_eq10}), and (\ref{supp_eq15}), the error signal can be simplified to
\begin{equation}
    i_{error}(\omega)=RP_0|T(\omega)|\times Sin(\psi(\omega)-\phi), \label{supp_eq16}
\end{equation}
where $|T(\omega)|$ and $\psi(\omega)$ are the amplitude and phase of the optical reference at the frequency of $\omega$. Please note that $\omega$ can be written as
\begin{equation}
    \omega(t) = \omega_0+\delta \omega_n(t), \label{supp_eq17}
\end{equation}
where $\omega_0$ and $\delta \omega_n(t)$ are the nominal frequency and frequency noise of the laser.\\
Supplementary Figure \ref{supp_fig_simulation_error}(b) and \ref{supp_fig_simulation_error}(c) show the numerical simulation of the frequency reference response and the asymmetric error signal, respectively. For numerical analysis purposes, we assume that the frequency reference is a critically coupled ring resonator with an approximately 80 GHz free-spectral range (FSR) (equivalent to a 1 mm circumference) and a Q-factor of around 3.3 million, corresponding to a propagation loss of approximately 0.2 dB/cm. As shown in Supplementary Figure \ref{supp_fig_simulation_error}(c), the error signal exhibits an asymmetric response around the frequency reference that can be used as a servo signal to lock the laser.

\newpage
\section*{Supplementary Note 3: comparative analysis of the PDH, an unbalanced MZI, and the cavity-coupled MZI}\label{supp_note_comparison}
\begin{figure}[b]
    \centering
    \includegraphics[width=\textwidth]{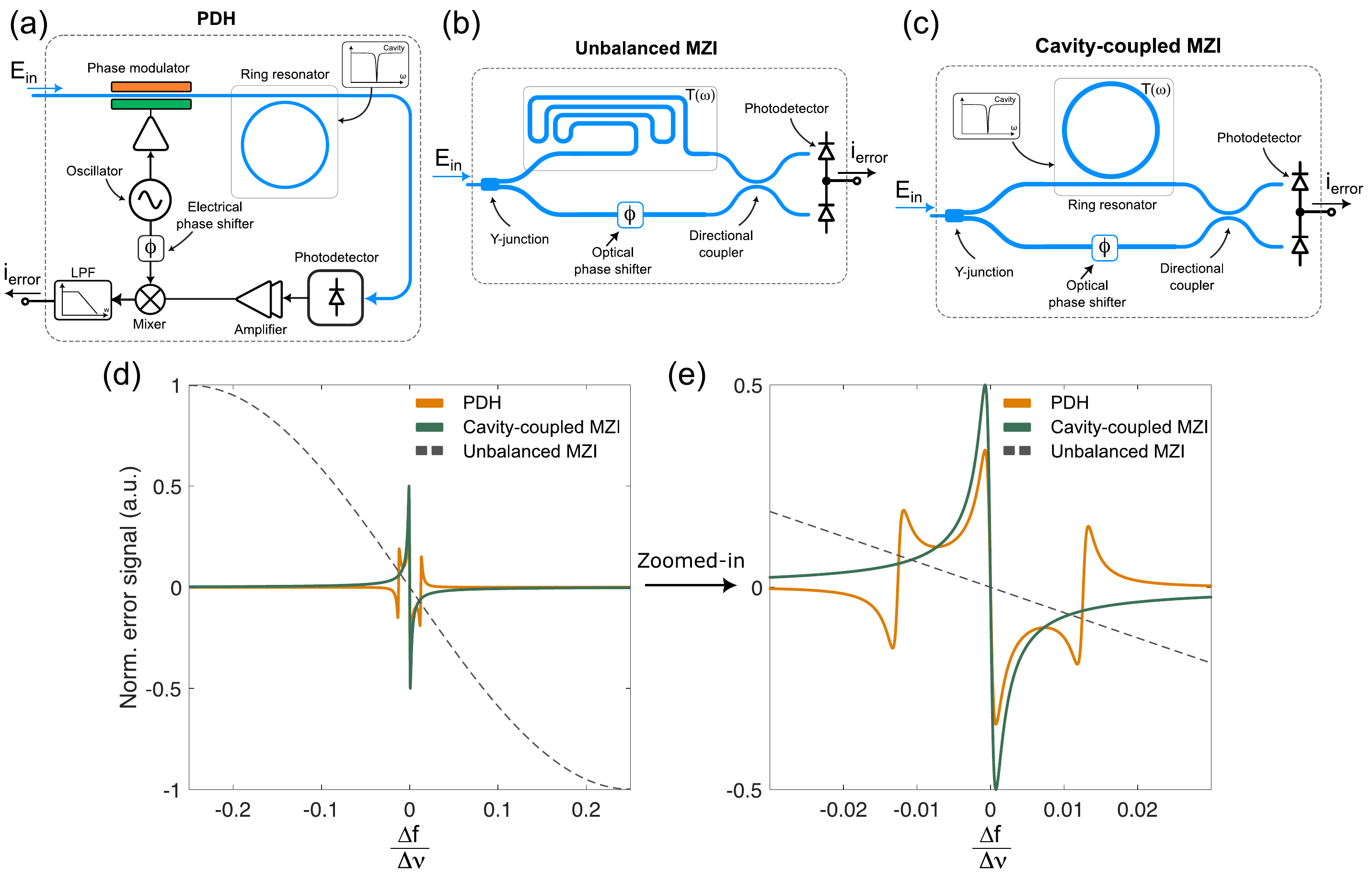}
    \caption{\textbf{Comparison between the PDH, an unbalanced MZI, and the proposed cavity-coupled MZI OFNDs.} (a) The block diagram of the PDH architecture. (b) The schematic of the unbalance MZI. (c) The schematic of the proposed cavity-coupled MZI. (d) and (e) are the simulated error signals of the aforementioned OFNDs. For this simulation, the circumference of the ring resonators in the PDH and cavity-coupled MZI schemes is identical to the waveguide length in conventional unbalanced MZI and is equal to 1 mm. The waveguide loss is assumed 0.2 dB/cm. $\omega, \omega_{ref}$, and $\Delta \nu$ are laser frequency, reference frequency, and the FSR of the ring resonator, respectively.}
    \label{supp_fig_comparison}
\end{figure}
Numerous OFNDs have been proposed and demonstrated in the past, including the well-known PDH architecture and the unbalanced MZI \cite{drever1983laser}. Supplementary Figures \ref{supp_fig_comparison} (a) to \ref{supp_fig_comparison} (c) show the block diagram of the PDH, an unbalanced MZI, and the proposed cavity-coupled MZI OFNDs, respectively \cite{idjadi2023enhanced}. As shown in Supplementary Figure \ref{supp_fig_comparison}(a), an electrical local oscillator modulates the incoming electric field using an optical phase modulator. The phase modulated signal is filtered by a cavity (optical frequency reference) followed by a photodetector. The photodetected signal is amplified using a trans-impedance amplifier and down-converted by the same local oscillator frequency using a frequency mixer. The down-converted signal is lowpass filtered to generate the error signal. The detailed mathematical derivation of the PDH error signal has been published previously \cite{black2001introduction}. \\
Supplementary Figure \ref{supp_fig_comparison}(b) shows the schematic of an unbalanced MZI where the frequency reference is the length difference (true-time delay) in one arm of the MZI. Using Supplementary Equation (\ref{supp_eq16}) and substituting $T(\omega)=e^{j\omega\tau}$, we can conclude that the error signal is sinusoidal. Supplementary Figure \ref{supp_fig_comparison}(c) shows the block diagram of the cavity-coupled MZI where the frequency reference is a microring resonator.  Supplementary Figure \ref{supp_fig_comparison}(d) and \ref{supp_fig_comparison}(e) display the calculated error signal and the zoomed-in view of it, respectively. To ensure a fair comparison, given the same chip area, the circumference of the ring in both the PDH scheme and the cavity-coupled MZI configuration is set to 1 mm, which is equivalent to the length mismatch in the unbalanced MZI arrangement. For the purpose of this analysis, a propagation loss of 0.2 dB/cm is assumed. Also, the local oscillator frequency in PDH architecture is 1 GHz, and the optical and electrical phase shifters ($\phi$) are optimized accordingly. As shown in Supplementary Figure \ref{supp_fig_comparison}(d), all three error signals are asymmetric around the reference frequency ($f_{ref}$). As suggested by Supplementary Figure \ref{supp_fig_comparison}(d), the proposed cavity-coupled configurations has significantly more sensitivity compared to a conventional unbalanced MZI ($K_{FD}=$8$\times$10$^{-11}$ Hz$^{-1}$). As plotted in Supplementary Figure \ref{supp_fig_comparison}(e), despite having a less complex design and potentially lower power consuming electronics, the proposed architecture exhibits a sensitivity ($K_{FD}=$1.7$\times$10$^{-8}$ Hz$^{-1}$) comparable to that of the PDH ($K_{FD}=$1.14$\times$10$^{-8}$ Hz$^{-1}$). It is important to highlight that in order to achieve the same OFND gain as the cavity-coupled MZI, an unbalanced MZI requires an equivalent delay of 17 nsec. However, implementing such a delay line, which is equivalent to approximately 1.27 m of TE-mode silicon waveguide, is not practical to implement on a silicon chip. Additionally, it avoids the need for phase modulation, which would introduce unwanted residual amplitude modulation \cite{shen2015systematic}. 

\section*{Supplementary Note 4: the OFND gain sensitivity analysis}
\begin{figure}[hbt]
    \centering
    \includegraphics[width=0.65\textwidth]{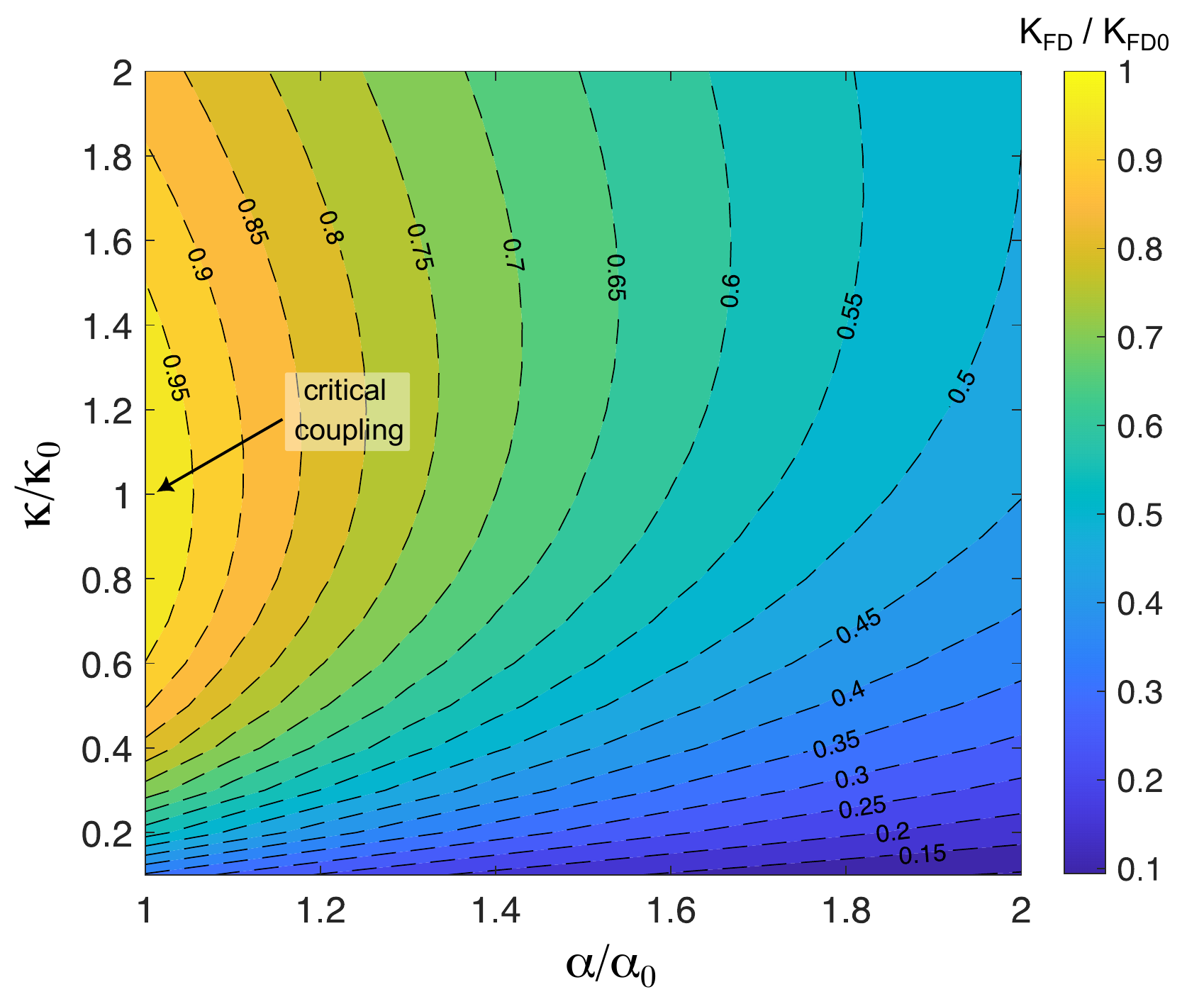}
    \caption{\textbf{The OFND gain sensitivity analysis to the ring resonator coupling and waveguide loss.} In this simulation, the OFND gain is normalized to the ideal scenario for a critically coupled ring resonator ($\kappa_0=0.5\%$) with circumference of 1 mm and waveguide loss $\alpha_0$ of 0.2 dB/cm. The maximum normalized gain ($K_{FD0}$) is 1.7$\times$10$^{-8}$ Hz$^{-1}$. }
    \label{supp_fig_sensitivity}
\end{figure}

As discussed in Supplementary Note 1, the gain of the OFND ($K_{FD}$) plays significant role in suppressing laser frequency noise. Therefore, careful design and parameter sensitivity analysis are necessary to ensure optimal performance. The main building block in the proposed cavity-coupled MZI OFND is the frequency reference which is an on-chip high Q-factor microring resonator. To achieve higher OFND gain, a microring resonator with higher frequency selectivity is essential. This can be accomplished by having a higher Q-factor and a large extinction ratio. The ring resonator coupling ratio ($\kappa$) and waveguide propagation loss ($\alpha$) are the two crucial parameters that ultimately affect the aforementioned ring properties and OFND gain.\\
Supplementary Figure \ref{supp_fig_sensitivity} shows the simulation of the normalized OFND gain sensitivity to the ring resonator coupling and waveguide loss. In this simulation, a ring resonator with circumference of 1 mm, waveguide loss ($\alpha_0$) of 0.2 dB/cm, and critical coupling condition ($\kappa_0=0.5\%$) is considered. Maximum gain of OFND ($K_{FD0}$) happens at the lowest waveguide loss ($\alpha/\alpha_0=1$) and critical coupling condition ($\kappa/\kappa_0=1$), which in this scenario is about 1.7$\times$10$^{-8}$ Hz$^{-1}$. Supplementary Figure \ref{supp_fig_sensitivity} demonstrates that for operating close to optimum performance (i.e., $K_{FD}/K_{FD0}>0.9$), it is necessary to have the waveguide loss within 10\% of the nominal value. However, the coupling ratio margin in this case does not require such a stringent constraint. It is worth mentioning that, even the fabricated chip may be in the lower performance regions because of the fabrication induced errors and inaccurate estimation of roughness, the overall loop gain can simply be compensated by the electronic amplifiers.
\newpage
\section*{Supplementary Table}

\begin{table}[hbt]
	\centering
  \begin{tabular}{|l|l|}
  	\hline
    \textbf{Equipment} & \textbf{Model} \\
    \hline
    DFB laser					&		AeroDiode 1550LD-2-0-0-1 \\
    Tunable laser						&		TOPTICA CTL 1550				\\
    Power supply					&		SIGLENT SPD3303X-E	\\
    Laser current source				&		LDX-3620B 							\\
    Laser temperature controller		&		LDT-5910B				\\
    Oscilloscope					&		SDS1204X-E	\\
    PID controller                  &       Moglabs FSC Servo Controller \\
    Optical spectrum analyzer		&		YOKOGAWA AQ6370D			\\
    \hline
  \end{tabular}
  \caption{List of equipment and components used in different experimental setups. }
\end{table}


